\documentclass[a4paper]{article}

\usepackage{booktabs}
\usepackage{threeparttable}  
\usepackage[backend=biber,citestyle=numeric]{biblatex}
\usepackage{rotating}
\usepackage{hyperref}
\immediate\write18{texcount -tex -sum  \jobname.tex > \jobname.wordcount.tex}

\providecommand{\keywords}[1]
{
  \small	
  \textbf{\textit{Keywords---}} #1
}

\title{A survey of multi-agent geosimulation methodologies: from ABM to LLM.}
 
\author{Virginia Padilla$^{1}$, Jacinto D\'avila$^{2}$ \\
        \small $^{1}${Departamento de Ciencia y Tecnolog\'ia}, \\
        \small {Universidad Nacional Experimental de Guayana}, \\
        \small {Estado Bol\'ivar}, {Venezuela.} \\
        \small $^{2}${CeSiMo, Facultad de Ingenier\'ia}, \\
        \small {Universidad de los Andes}, \\
        \small {Estado M\'erida}, {Venezuela.}  \\
} 
\date{} 

\begin{document}
\maketitle

\begin{abstract}
	{We provide a comprehensive examination of agent-based approaches that codify the principles and linkages underlying multi-agent systems, simulations, and information systems. Based on two decades of study, this paper confirms a framework intended as a formal specification for geosimulation platforms. Our findings show that large language models (LLMs) can be effectively incorporated as agent components if they follow a structured architecture specific to fundamental agent activities such as perception, memory, planning, and action. This integration is precisely consistent with the architecture that we formalize, providing a solid platform for next-generation geosimulation systems.} 
\end{abstract} \hspace{10pt}

 
\section{Introduction}\label{sec1}

This paper presents a review of the literature on agents and geosimulation. Drawing from projects that have modeled and implemented agents for at least two decades, we extracted evidence to validate a framework serving as a formal specification for a geosimulation platform. 
The need for formalizations of complex systems is long-standing. The relative weakness of software engineering in the field of multiagent systems, for instance, has been acknowledged: "There are numerous methodologies or object-oriented languages available, but no firm commitment to a specific operational semantics" (Drogoul, 2003). With the advent of new forms of artificial intelligence (AI), such as large language models (LLMs), this need has grown. 

Our goal is to provide a more comprehensive, consistent, robust, and dependable platform for knowledge management services by conceptualizing, building, and organizing agent concepts into multi-agent systems (MAS).  To this purpose, we propose a conceptual framework for agents that can be compared to the agent characterizations provided by multi-agent system development approaches. 

It is then combined with a reference model that attempts to explain the interactions between agents, databases, and geographic information systems as tools for modeling and simulating complex geographical systems. Then we undertake a thorough bibliographic review to have a better understanding of how geosimulation components were produced. 

\section{Agents.} \label{sec2}
The notion of agent has become extremely popular in the technological world in recent times. AI revolves around this concept \cite{Russell2004} and is aiming to develop a new agent-oriented paradigm to reinforce the object-oriented paradigm \cite{Rumbaugh1991}. According to Russell \cite{Russell2004} ,  an agent is an object with an interface to its environment through which inputs arrive and outputs are produced.

The internal dynamics of an agent, which affects its internal states and strongly links its inputs and outputs, are what distinguish it as an active and unique object. When this connection produces a type of behavior, then Russell \cite{Russell2004} talks about intelligent agents. It can be said that the goal of the entire AI project is to define certain types of behavior and find a way to generate them.

To make an agent an object of a specific type, the agent-oriented paradigm~\cite{Kowalski1987, Bratman1987, Russell2004, Shoham1991} prescribes a set of structures for that internal state. What follows is the specification of a multiagent system constituted by rational agents, described by their internal substructures, but also including their interfacing and operations contexts~\cite{Russell2004, Bradshaw1997, Davila1997}. That structure constitutes a model that has been formalized by \cite{Kowalski1997, Kowalski1999, Kowalski2006, Kowalski2009, Kowalski2011, Davila1997, padilla2013a}.

\subsection{The Agent Reference Model.}
\label{ARM}

The \textbf{ Agent Reference Model}, ARM, is a conceptual framework in which we describe the elements of an agent with a view to implementations. It has been formalized in \cite{Padilla2012, padilla2013a} and is used (in Section~\ref{methodologies}) as a comparative framework to evaluate industrial methodologies, showing how or whether each of its concepts is represented. The goal (beyond the scope of this paper) is to approach a meta-methodology, based on \cite{Davila2007}, \cite{Davila2005}, \cite{Davila2005a}, \cite{Davila2005b}, \cite{Davila2002}, \cite{Davila2002a}, \cite{Davila2000}, and \cite{Davila1997}, which allows exploring concrete strategies to design, generate, and control multi-agent systems. This is the Agent Reference Model (ARM):
\begin{itemize}
	\item Internal state structures:
	\begin{itemize}
		\item{Beliefs}: what the agent knows about its environment and other agents.
		\item{Goals}: objectives or situations that the agent or its designer would like to achieve or cause, usually through the execution of a plan. These can be classified into
		\begin{itemize}
			\item Maintenance goals, which represent a permanent relationship between the agent and its environment, in the form of conditional rules.
			\item Achievement goals are particular objectives that the agent tries to achieve at some point in time.
		\end{itemize}
		\item{Intentions}: goals pre-selected for reduction and action execution.
		\item{Preferences}: a distinguished set of goals. An agent's preferences for a certain state can be incorporated as part of a utility function, to which values are assigned to express how desirable each state or goal is.
		\item{Commitments}: the obligations (transformed into the agent's goals) acquired or agreed upon with other agents and to which the agent is subject.
		\item{Plans}: the sequences of actions that an agent can execute to achieve its goals.
		\item{History}: the agent stores information concerning its record of perceptions and actions.
	\end{itemize}
	\item{Internal dynamics}:
	\begin{itemize}
		\item Knowledge and beliefs updating mechanism.
		\item Agent activation mechanism.
		\item Agent planning and execution mechanism, which includes an inference engine and a decision-making mechanism.
	\end{itemize}
	\item{External State}:
	\begin{itemize}
		\item{Roles}: organizational functions performed by the agent in a multi-agent system. They are generally represented by goals.
		\item{Use cases}: description of the agent's behavior.
	\end{itemize}
	\item{Interface}:
	\begin{itemize}
		\item{Skills}: The agent has the correct functionality and information to be able to interact with the environment surrounding it. It is defined by two attributes:
		\begin{itemize}
			\item{Abilities}: what the agent can do as a response to the combination of his perceptions and beliefs.
			\item{Capabilities}: the set of actions that an agent can perform under certain preconditions provided.
		\end{itemize}
	\end{itemize}
\end{itemize}

\subsection{Methodologies for the development of multi-agent systems.}
\label{methodologies}
Table \ref{tab-ARM} shows a set of recognized MAS methodologies versus the descriptor concepts of an agent, as described in the ARM. Methodologies under scrutiny are:  {AAII} methodology \cite{Rao1995a} \cite{Rao1995} \cite{Rao1991}, GAIA, a methodology with a high level of abstraction \cite{Zambonelli2003},  {MaSE}  \cite{Mark2000},  Prometheus, as proposed by \cite{Padgham2002}, {MESSAGE/UML}, which appeared in \cite{Caire2002}, {INGENIAS}, as proposed by \cite{GomezSanz2003},  Tropos \cite{Bresciani2004}, {MAS-CommonKADS}    \cite{Iglesias1996}, and {O-MaSE} \cite{GarciaOjeda2007}. 

	\begin{sidewaystable}[!t]%
		\centering %
		\caption{The Agent Model Reference vs Multi-Agent  methodologies
			\label{tab-ARM}}%
		\tiny
		\begin{tabular*}{0.5\textwidth}{@{\extracolsep\fill}llllllllll@{\extracolsep\fill}}
			\cmidrule{1-10}
			\textbf{ARM} & \textbf{AAII}  & \textbf{GAIA}  & \textbf{MASE}  & \textbf{Prometheus} & \textbf{MESSAGE/UML} & \textbf{INGENIAS}  & \textbf{Tropos}  & \textbf{MAS-CommonsKADS}  & \textbf{O-MaSE}\\
			\cmidrule{1-10}
			Internal state  &   &    &    & &   &   &    &    &    \\
			Beliefs & X  &  \tnote [$^\dagger$]  & X  & X & X & X  & X  & X  & X  \\
			Goals & X  &     & X  &   & X & X  & X  & X  & X  \\
			maintenance &    &     &   &   &   &   &   &   &   \\
			achievement &    &   X  &   &    &  X  &   &   &   &   \\
			Intention   & X   &     & X  &  X & X  &  X & X  & X  &  X \\
			Preference  &    &     &   &   &  X &   & X  & X  &   \\
			Commitments &    &   X  &   & X  &   & X  & X  & X  &   \\
			Plan & X   &     & X  &  X &   & X  &  X &   &   \\
			maintenance &    &     &   &   &   &   &   &   &   \\
			History &    &     &   &   &   & X  &   &  X &   \\
			\cmidrule{1-10}
			Internal Dynamics  &   &    &    &   &   &   &    &    &    \\
			Update  mechanism &   &     &    &   &  &    &   &  X &    \\
			activation mechanism &   &     &    &   &  &    &   & X  &    \\
			planning mechanism &   &     &    &   &  &    & X  & X  &    \\
			\cmidrule{1-10}
			External State &   &     &    &   &  &    &   &   &    \\
			Roles &  X &  X   &  X  & X  &X  &  X  & X  & X  &  X  \\
			Use case      &  X &   X  &  X  &  X & X & X   &   &  X &  X   \\
			\cmidrule{1-10}
			Interface &   &     &    &   &  &    &   &   &    \\
			Skills   &   &     &    &   &  &    &   &   &    \\
			abilities &   &  X   &    &   &  &    &   &   &    \\
			capabilities &   & X    &    &  X &  &    & X  &   &    \\
			\cmidrule{1-10}
		\end{tabular*}\hspace*{350pt}
		\begin{tablenotes}
			\item  $^\dagger$GAIA does not provide details on its internal architecture.
		\end{tablenotes}
	\end{sidewaystable}

As can be seen , no methodology includes all the concepts in the ARM. Instead, the ARM does include all the features and concepts in the listed methodologies. This is the reason we believe that the ARM is a general model of an agent that encompasses the different visions of MAS offered by the reviewed methodologies.  

In the following section, we focus the review on the area of geosimulation, to understand the models and technologies that have been used to model complex systems with geographical dynamics.

As can be seen, no methodology includes all the concepts in the ARM. Instead, the ARM does include all the features and concepts in the listed methodologies. This is the reason we believe that the ARM is a general model of an agent that encompasses the different visions of MAS offered by the reviewed methodologies.  

In the following section, we focus the review on the area of geosimulation, to understand the models and technologies that have been used to model complex systems with geographical dynamics.

\section{Geosimulation and Agent-Based Models} \label{sec3}
Benenson \cite{Benenson2004} further identifies geosimulation as a new area of study and a chance to help create new tools by defining it as the fusion of three technologies: 1) modeling and simulation, 2) software agents, and 3) a geographical information system, or GIS. For his part, Blecic \cite{Blecic2009} has said that multiagent geosimulation is a simulation technique to model phenomena that occur in geographical areas using an agent-based approach in high-resolution spatial models. 
This section reviews geosimulation models and tools for complex systems with geographic dynamics. First, we provide a generic formalization that describes the links between an agent's formal model and a database model in a multi-agent system. Then, we conduct an exhaustive bibliographic review to better understand how the components of a geosimulation were developed. 
The research works have been classified according to whether their development has been based on (a) a cognitive framework that has used a cognitive theory to model agents, (b) a generic framework with constructs from software engineering, (c) generic geosimulation tools those that use geosimulation's platforms to implement the solution, and (d) generative agents.  

In the following section, we introduced a general formalization where we explain the relations between the formal model of an agent and a database model in a multi-agent system.

\subsection{A new approximation to a formal and embodied model of a multi-agent system}
A reference model is the epistemic basis of an ontology. It is the description of objects and concepts within some domain of knowledge\footnote{ontology in its historical sense, rather than the modern, technical meaning.}. In what follows, we propose a mathematical specification of a multiagent system to guide computational implementations of such kinds of system, particularly those applied to knowledge management and simulation. We are building on previous work in AI \cite{Ferber1996} and Simulation\cite{Davila2005, Davila2005a} which led to a multi-agent theory for simulation\cite{Davila2007} which guided the development of the GALATEA simulator.
GALATEA is a simulation software that integrates, in the same computational platform, the conceptual and concrete tools for the simulation of discrete events, continuous systems, and multiagent systems; in a distributed and interactive way.  The simulation software is based on the general formalism of modeling and simulation of discrete events, DEVS \cite{Zeigler2000}, \cite{Wainer2009}, and on the theory of Simulation of Multi-Agent Systems that can be consulted in \cite{Davila2007}

Here we adapt that theory and connect it to the MAGI theory by \cite{Blecic2009} to produce a multi-agent theory for geosimulation, which seeks to explain the relations between agents, databases, and geographic information systems as tools to model and simulate complex spatial systems. For the sake of clarity, we start by reproducing the MAGI metamodel.

\label{MAGI-theory}
The MAGI theory\cite{Blecic2009} is a metamodel that amounts to a formal theory of geography with agents and objects in it. This MAGI theory is a perfect complement for our multiagent theory as it provides for 1) the embodiment of agents and 2) a carefully tailored account of the data structures and associated functions required for a geographic information system to efficiently compute answers to queries. A third side effect of the combination of these theories is the possibility of accounting for the creation of objects and agents. The theory of Galatea did not have those elements. On the other hand, the combination of MAGI and GALATEA provides an explicit account of time and a DEVS strategy for time management.

In MAGI theory, the environment, $ Env $, is characterized by a 3-tuple from the cross product of 1) the set of all possible global $ parameter=value $ pairs to describe a system, 2) the set of all possible global functions operating on those parameters, and 3) the set of all possible layers, $ L $, of objects that may constitute the geography of the system. Each layer, $ L \; \epsilon \; \mathcal{L} $, is characterized in turn, but another 3-tuple from 1) the set of all possible local parameters, 2) the set of all possible local functions, and 3) the set of entities (objects and agents) that populate the system.

Agents are, in turn, described by a double record: the agent itself and its type $\tau$. The agent is described by its internal state, its geo-spatial attributes, and the set of references to objects and other agents observed by this agent and the subjects of its actions. An agent type $\tau$ is described by a 6-tuple: 1) the set of all possible internal states of the agent, 2) the set of admissible shapes for this type of agent, 3) the set of all possible actions, 4) the set of perception functions, 5) the set of decision functions, and 6) the set of agreement functions by which this agent cooperates with other agents. It should be clear that these correspond to an embodied account of a multi-agent system because these agents have well-defined attributes for their bodies and locations in a physical space.

All these elements are formalized in the combined theory
~\cite{padilla2013a}, which constitutes a formal description of each agent and its environment in a multi-agent system. Examples of systems that fit to this description are presented in the following section. 

\subsection{Cognitive Frameworks}
\paragraph{BDI's Agents:}
Vahidnia \cite{Vahidnia2015} uses BDI's agents (BDI: Beliefs, Desires, Intentions) to describe and develop an architecture that combines a multi-agent system with GIS, logical deduction, and qualitative reasoning. The system integrates multiple moving agents and the concept of means-ends spatio-motional reasoning. The architecture has a complementary quantitative component that supports collaborative planning based on the concept of equilibrium and game theory.

\paragraph{Possibilistic BDI:}
Possibilistic BDI agents were presented by Costa Pereira \cite{CostaPereira2010} and his description states an agent with internal mental state $S$ that is described by a possibility distribution $\pi$, representing beliefs, and by a set of desire-generation rules $RJ$. Possibility distribution $\pi$ is dynamic and changes as new information $\phi$ is received from a source.
The agent rationally elects its goals $G$ $*$ from the justiﬁed desires $J$ as the most desirable of the possible sets of justiﬁed desires, according to a possibility measure $\Pi$ induced by $\phi$. The agent then plans its actions to achieve the elected goals $G$ $*$ through a planner module.
Vanegas Hernandez et. al. \cite{VanegasHernandez2017} have studied the feasibility of using BDI agents for modeling the phenomena of urban growth and segregation. For this purpose, they have introduced a framework that allows to representation of households, investors, and promoters, while using possibilistic BDI agents \cite{CostaPereira2010}, interacting over a spatial context as a segregation model proposed by Schelling \cite{Schelling1971}.
\paragraph{CAUSE:}
Wozniack \cite{Wozniak2020} has developed a universal conceptual framework for building agent-based models of real cities: Complex Artificial Urban Systems (CAUSE). The geographical space in CAUSE is projected through GIS data. Agents are described through the framework of Maslow's pyramid \cite{Maslow1943}. His focus on the labor market and real estate market are modeled through the agent-based matching function approach. Inhabitants of the artificial city try to achieve the highest possible level of satisfaction maximizing individual utility functions.


\subsection{Generic Frameworks for Agent-Based Simulation}
\paragraph{MAGS:}
MAGS, a platform developed by \cite{Moulin2003}, is a generic software platform for the creation of Multi-Agent Geo-Simulations involving several thousand agents interacting in virtual geographic environments (in 2D and 3D). This platform was used for the simulation of crowd behaviors in urban environments.
MAGS agents possess cognitive spatial activities:
\begin{enumerate}
	\item an agent should be able to perceive the spatial environment, objects, and other agents;
	\item a GIS and related databases are the core to generate the spatial environment including the static objects;
	\item the agents are described with internal state and goals, and the capacity to plan their activities according to the information they perceive in the virtual space.
\end{enumerate}

MAGS was used as a basis to present several simulation platforms that model different application areas such as Train-MAGS \cite{Sahli2008}, an agent-based geosimulation tool that simulates train behaviors and identifies risky areas in large-scale geographic environments.
Mekni and Moulin \cite{Mekni2008} have used the MAGS platform to develop a multi-agent geosimulation approach to analyze and manage sensor networks that are deployed in large-scale geographic environments for in situ sensing and data acquisition purposes. This approach has been applied in the context of a water resources monitoring project.
In Ekemas \cite{Sahli2009}, an agent-based geosimulation framework assists human planners when planning under strong spatial constraints in a real large-scale space. The approach consists of drawing a parallel between the real environment and the simulated environment based on GIS data. This virtual environment uses software agents that are aware of the space and equipped with advanced spatial reasoning capabilities. Bouden and Moulin \cite{Bouden2010} have proposed an extension to MAGS, called ZoonosisMAGS, a geosimulation tool to simulate the propagation of the West Nile Virus.

Haddad and Moulin \cite{Haddad2010} presented a framework based on a conceptual model of spatio-temporal situations along with MAGS. The framework was able to propose courses of action that in order to change towards a more realistic geographic space. In terms of reasoning, Haddad and Moulin (.ibid) identify causation relationships between spatio-temporal situations of historical events relating to an agent.

Mekni and Moulin \cite{Mekni2011} have also proposed an approach that extends another Informed Virtual Geographic Environment (IVGE) model \cite{Mekni2010b} to manage knowledge of the environment and support agents' cognitive capabilities and spatial behaviors. This approach relies on previous well-established theories on human spatial behaviors and the way people apprehend the spatial characteristics of their surroundings to navigate and interact with the physical world. It is also inspired by Gibson's work \cite{Gibson1979} on affordances and knowledge provided by the environment to guide agent-environment interactions.

Haddad \cite{Haddad2019} have proposed modeling and analyzing the risk of workers' exposure to hazards in a port environment; a spatial and temporal problem, given that safety risks are often closely related to the proximity of workers to nearby hazards. The MAGS platform has been used to model the dynamic environment of a port. A multi-agent system is used to model, in the Virtual Geographic Environment (VGE), the behavior of real entities and actors of the real world (workers, trucks, heavy machines, etc.) and to track their movements.
\paragraph{PARKAGENT:}
Benenson and Master \cite{Benenson2008} have proposed PARKAGENT, an agent-based, spatially explicit, model for parking in the city. PARKAGENT is based on the geosimulation approach, combining a real-world ArcGIS database with a multi-agent system. The model simulates the behavior of each driver in a spatially explicit environment and can capture the complex self-organizing dynamics of a large collective of parking agents within a non-homogeneous (road) space.
\paragraph{COLMAS:}
The COLMAS Project, presented by Perron \cite{Perron2008}, aims at developing a framework, algorithms, and automated advisory decision support capabilities for dynamic distributed resource management in which a heterogeneous team of agents drawn from distinct classes (static and moving airborne/land vehicles, unmanned/manned vehicles) while is engaged in a surveillance mission (reconnaissance, target search including detection/ recognition, information gathering, exploration, etc.) evolving in a dynamic uncertain environment with both known and unknown targets and threats (a mix of moving/static, evading/non-evading behaviors).
\paragraph{MAGI:}
As explained in section~\ref{MAGI-theory}, The MAGI theory \cite{Blecic2009} is a metamodel that is equivalent to a formal theory of geography with agents and objects in it.
\paragraph{Metro{N}et:}
Blumenfeld \cite{BlumenfeldLieberthal2012} have developed MetroNet, a USM (Urban Simulation Model) specifically designed to study the evolution and dynamics of systems of cities. Its structure is a superposition of cellular automata and agent-based modeling approaches (spatial analysis) and a complex network approach (topological analysis). The agents in its model represent workers who look for working places. This work aims to identify a set of fundamental rules that govern the interactions within urban systems at the metropolitan scale.
\paragraph{SIENA:}
Fetch \cite{Fecht2014} has proposed SIENA, an urban simulation model for environmental health analysis, a tool to explore urban interactions and processes about exposure assessments. The development of SIENA involved identifying and quantifying fundamental processes and similarities in urban areas and using those to guide the building of SIENA within a GIS. SIENA supports probabilistic models in the formulation of laws that control the behavior of the system.
\paragraph{ReHoSh:}
Rienow and Stenger \cite{Rienow2014} have combined the urban cellular automaton SLEUTH and a multi-agent system, ReHoSh (Residential Mobility and the Housing Market of Shrinking City Systems), to simulate residential mobility in a shrinking city agglomeration: residential mobility and the housing market of shrinking city systems focuses on the dynamic of interregional housing markets implying the development of potential dwelling areas. An agent is defined as an abstract entity (for example, a home or community) that is autonomous, intelligent, mobile, and adaptive.
\paragraph{Emergency of riots:}
The emergence of riots has been categorized by Torrens and McDaniel \cite{Torrens2013a} as a complex system. To capture this complexity, Pires, and Crooks \cite{Pires2017} have developed a theoretically grounded agent-based model (ABM) that integrates ABM with geographic information systems (GIS) and social network analysis (SNA), through the lens of geosimulation, to explore how the environment and local interactions at Kibera (an informal settlement located within Nairobi), combined with an external trigger, such as a rumor, led to the emergence of riots.
Agents have been modeling using the PECS (Physical conditions, Emotional state, Cognitive capabilities, and Social status) framework \cite{Kennedy2012}. Furthermore, they delineate the agents with the theory of the human hierarchy of needs proposed by Maslow \cite{Maslow1943}
\paragraph{MATSim:}
Ben-Dor et al \cite{BenDor2020} has developed the Multi-Agent Transportation Simulation (MATSim). It is an agent-based traffic model that includes intrinsic down-scaling: procedures of changing network parameters to simulate the dynamics of the system as a whole while activating only a fraction of travelers.
\paragraph{Genetic Algorithms:}
Fu \cite{Fu2020} has proposed the problem of route planning for security patrol in smart communities, a simulation framework comprising of multi-agent-based model and genetic algorithm (GA). The GA is used to determine and evolve the route collection and find the optimal results, while the multi-agent simulation model can be used to set constraints and get the objective values of routes. The implementation of the simulation system is based on Anylogic, which is beneficial for interacting with the GA program code.

\subsection{Generic Software Plataforms}
\paragraph{NETLOGO:}
Fisher and Lassa \cite{Fisher2017} have used the NetLogo ABM platform to produce a spatially explicit cellular automata model and a geosimulation ABM to model complex environmental social and political considerations. It was to be incorporated and visualized in the domain of Travel and health service area, developed to assess access to emergency obstetric care in rural areas.
\paragraph{REPAST:}
Dragićević and Hatch \cite{Dragicevic2018} have implemented the Logic Scoring of Preference (LSP) as a method to represent the human decision-making process of agents in an ABM of land-use change. The proposed LSP-ABM method simulates residential land-use change at the cadastral level. Various stakeholder types including residents, developers, and city planners are integrated as agents in the geosimulation model.
\paragraph{GeoMason:}
Kim et. al. \cite{Kim2019a} have assigned the term geo-social to models to simulate individuals (i.e., agents) with plausible social behavior that is based on Maslow's psychological and social science theories \cite{Maslow1943}.
Geo-social agents were used \cite{Kim2020} in a framework with GeoMASON \cite{Sullivan2010}, and its GIS extension, adding a disease model that simulates an outbreak and allows testing different policy measures such as implementing mandatory mask use and various social distancing measures. GeoMASON is a extension of the MASON (Multi-Agent Simulation of Neighborhoods) open-source simulation toolkit \cite{Luke2005}.
\paragraph{GeoMason and Jade:}
Z$\ddot{u}$efle et. al. \cite{Zuefle2020} have introduced the Urban Life agent-based simulation used by the Ground Truth program to capture the innate needs of a human-like population and explore how such needs shape social constructs. This model was used to predict future states and to prescribe changes to the simulation to achieve desired outcomes in a simulated world.
Z$\ddot{u}$efle's Agents follow a pattern of life with a every day cycle based on the augmentation of Maslow's hierarchy of needs \cite{Maslow1943}. To support large-scale urban life simulations, the authors have designed a framework by integrating the multi-agent systems toolkit JADE \cite{Bergenti2014} with GeoMASON \cite{Sullivan2010}.
\paragraph{GAMA:}
Macatulad and Blanco \cite{Macatulad2014} have developed a multi-agent geosimulation model for evacuation of buildings, integrating the 3D-GIS dataset of the case study of buildings as input in an ABM using the GAMA simulation platform.
In another document, Macatulad and Blanco \cite{Macatulad2019} have enlarged the previous study with the developed a three-dimensional geographic information system (3D-GIS)-based multi-agent geo-simulation model was developed using the GAMA simulation platform integrating 3D-GIS layers and agent-based modeling for evacuation scenario modeling.
Bandyopadhyay and Singh \cite{Bandyopadhyay2018} have proposed a microsimulation approach based on agent-based modeling and they consider the spatial aspects of urban areas to recreate the dynamic emergency environment for assessment of urban emergency response plan (UERP).
The model was implemented in GAMA 1.5.1 (GIS Agent Modelling Architecture), a spatially explicit multi-agent modeling platform.
GAMA provides integration of GIS and agent modeling capability \cite{Taillandier2012}. Agents are defined following principles of the BDI architecture \cite{Wooldridge1998}.
\paragraph{PNM:}
Rimbaut \cite{Raimbault2020} put forward the basis of an agent-based model that can eventually evaluate theoretical optimal symbiotic exchanges given a set of actors in a geographical area and compare it to empirical or alternative scenarios. For this objective, they have developed a spatial ABM in which agents are industrial companies. In the model of industrial symbiosis, they have replaced the one-dimensional niche space from the probabilistic niche model (PNM) \cite{Williams2011}, which can reproduce the structure of complex food webs, with an one dimensional 'by-product' space along which the input and output functions for each company that are deﬁned as Gaussians.
\paragraph{Mesa:}
In Gharakhanlou and Perez \cite{MahdizadehGharakhanlou2022} the purposes of this study were to develop a spatially explicit agent-based model to simulate the dynamics of COVID-19 spread and assess the effectiveness of two control interventions in containing the COVID-19 outbreak in the city of Montreal, QC, Canada.	The simulation of the COVID-19 outbreak in this study was implemented in the Mesa framework (i.e., an ABM framework in Python) \cite{masad2015}.
\paragraph{Python-based ABM framework:}
Yin et al.\cite{10.1145/3681770.3698571} demonstrate the possibility of scaling an agent-based model for a massive population (20 million humans in the State of NY) to consider the effects on individuals of particular attributes such as social influence. They use a geosimulator with spatial information in more than one layer, namely physical space (real geography), cyberspace (like social networks), and working space relationships between agents, to characterize individual attitudes towards vaccination.
These are practical developments of massive, spatially aware simulations that have been formalized before, as explained in the following section.  

\subsection{Geographic Automata Systems, GAS}
Torrens and Benenson \cite{Torrens2005} have designed an approach to specify simulated geographic objects as geographic automata that combine CA and MAS concepts in unique ways, by considering collections of interacting geographic objects as Geographic Automata Systems. 
This framework takes advantage of the formalism of automata theory and Geographic Information Science to combine cellular automata and multi-agent systems techniques and provides a spatial approach for bottom-up modeling of complex geographic systems that are comprised of infrastructure and human objects.

Formally, a Geographic Automata System (GAS), $G$, may be defined as consisting of seven components: $G = (K; S, T_{s}; L, M_{L}; N, R_{N})$ where $K$ is the set of types of automata featured in the GAS, $S$ is the set of states and $ T_{s}: (S_{t}, L_{t}, N_{t} ) \rightarrow S_{t+1}$ is the set of state transition rules, used to determine how automata states should change over time. 
$L$ contains the georeferencing conventions that dictate the location of automata in the system, $ M_{L}: (S_{t}, L_{t}, N_{t} ) \rightarrow L_{t}+1 $ are the movement rules for automata, governing changes in their location in time, $N$ represents the neighbors of the automata and $R_{N}$ represents the neighborhood rules that govern how automata relate to the other automata in their vicinity. $ R_{N}: (S_{t}, L_{t}, N_{t} ) \rightarrow N_{t+1}$ specifies this condition. GAS has been used in the geosimulation of the complex urban phenomenon described in the following works: 

Sabri et al., \cite{Sabri2012}, have used GAS to design a conceptual framework for geosimulation of the New-build gentrification process in an integrated approach, where	the combination of Multi-Criteria Evaluation (MCE) and GAS facilitates to translate the expert Knowledge into model rules.
Cabs et al. \cite{Cabs2013} in the study on pedestrian behavior in spatial environments, data collection, and simulation methods of pedestrian movement models.
Torrens et al. \cite{Torrens2013} have presented an architecture to achieve simple and complicated realizations of urban sprawl in simulation. GAS has been used to represent the geographical drivers of sprawl in intricate detail and over fine resolutions of space and time in the context of the messily complex and complicated urban processes and phenomena that work within city sprawl geography.
Torrens and McDaniel \cite{Torrens2013a} have introduced poly-spatial agents, based on socio-emotional agents \cite{Epstein2002}, with the ability to adapt their behavioral geography under changing circumstances and to process geographic information from diverse sources. It is an approach to modeling riot-prone and riotous crowds using behavior-driven computational agents.

Torrens et al. \cite{Torrens2015a} have set GAS and polyspatial automata, a wrapper around GA that serves to control the nature of the set of state transition rules, particularly as they relate to space-time scale and the context it provides.
Torrens and Gu \cite{Torrens2021} had adapted GAS to a model agent with variations in their set of featured automata and state transition rules, with the purpose that geosimulation can be used in close connection with virtual geographic environments and virtual reality environments to build human-in-the-loop interactivity between real people and geosimulation of the geographies that they experience. 

The work of Torrens and Kim \cite{Torrens02012024} is hard to circumscribe, as it involves multiple devices in a real space organized to embed humans in simulations of real-life situations, such as road crossing. Its objective is to obtain information on the fidelity and similarity of the embedded experiences.

\subsection{Generative Agents: LLM as agent-components}
The work at Stanford by Park et al.\cite{park2023generativeagentsinteractivesimulacra} is a seminal and systematic effort to embed Large Language Models within agents, which they termed as Generative Agents, in simulations. It is particularly striking to see how an LLM fits as the perception mechanism of an agent by reading the state of the world from a description of it in natural language. The report is then stored in a memory reserved for that agent, so that it keeps a trace of the changes in the world. Those changes are then retrieved and fed back to an LLM, this time with a suitable prompt to ask the LLM to produce a plan for the agent to go about. So, the LLM is also used as the reasoning mechanism and, more importantly, as the source of knowledge about how to behave in the world. The authors demonstrate the potential of generative agents in a Sims-style game world and then evaluate their behavior to conclude that this architecture creates believable behavior in human social situations.

Xi et al. \cite{xi2023risepotentiallargelanguage} postulate that large language models (LLMs) represent a promising avenue towards artificial general intelligence, due to their emerging capabilities in knowledge acquisition, reasoning, planning, and natural language understanding.
They proposed a framework for LLM-based agents consisting of three modular components: the brain, perception, and action. The article analyzes the use of LLM-based agents in a simulated environment to study social phenomena. Highlight notable social experiments such as the Hawthorne experiment and the Stanford Prison experiment and propose the idea of a "society of agents" where the behaviors and personalities of these agents are analyzed. It shows how agents exhibit complex and emergent social behaviors influenced by cognitive processes and environmental factors. The article also categorizes these behaviors into key dimensions such as information absorption, internal cognitive processing, and social behavior. Similarly to the work at Stanford by Park et al. mentioned in the previous paragraph, the authors conclude that LLMs can be used as components of agents, provided that they conform to an architecture and representation as required for each component: Perception, Memory, Planning, and Action. Exactly the kind of architecture we have been formalizing. Another comprehensive review by Li et al~\cite{li-et-al2024} refers to more than 20 examples of multi-agent systems using LLMs as agent components in approximately the same structures.

\section{Discussion} \label{sec4}
Having introduced a general formalization that explains what an agent and a multi-agent system are, and reviewed that wide range of applications and tools for multi-agent simulations, we are in a position to briefly explain the critical components of a multi-agent system, MAS\cite{Davila2007, Ferber1996, Padilla2013}.  

A MAS requires agent as computational objects, each one with a well-structured internal state and specific dynamics to update that state from time to time or event to event. That internal state can be characterized as a structured memory for the agent's beliefs, its desires, goals, or intentions, and records of what the agent perceives from and does upon its environment. The agent also requires specific methods to perceive and act, but more importantly, to meaningfully connect perceptions with actions, by some systematic process of reasoning and planning to achieve its goals. 

Apart from the internal state and dynamics of the agents, a MAS must also provide for some interfaces between agents and their environment, so that the agents can perceive, according to their actual position and perceptual abilities, and act, according to their capabilities. And, of course, a MAS can also provide support for phenomena independent from the agents and more depending on natural, geographical or spatial dynamics. 

\section{Conclusion and Future Work} \label{sec5}
We have surveyed agent-based methodologies that formalize the concepts and relations between agents in multi-agent systems, simulations, and information systems. We compared them using a formal model, ARM, that establishes the bases for a common language (with diverse syntactic and graphical expressions) with extended semantics to integrate rules, events, time management, conditions of operation, and other database constructs into the agent-oriented paradigm. We offer this formal model as a specification to implement a geosimulation system able to integrate, faithfully model and simulate systems with intelligent agents.  

\printbibliography

@Article{li-et-al2024,
  author    = {Xinyi Li and Sai Wang and Siqi Zeng and Yu Wu and Yi Yang},
  journal   = {Vicinagearth},
  title     = {A survey on LLM‑based multi‑agent systems:workflow, infrastructure, and challenges},
  year      = {2024},
  number    = {9},
  volume    = {1},
  doi       = {10.1007/s44336-024-00009-2},
  publisher = {Springer},
  url       = {https://doi.org/10.1007/s44336-024-00009-2},
}

@misc{park2023generativeagentsinteractivesimulacra,
      title={Generative Agents: Interactive Simulacra of Human Behavior}, 
      author={Joon Sung Park and Joseph C. O'Brien and Carrie J. Cai and Meredith Ringel Morris and Percy Liang and Michael S. Bernstein},
      year={2023},
      eprint={2304.03442},
      archivePrefix={arXiv},
      primaryClass={cs.HC},
      url={https://arxiv.org/abs/2304.03442}, 
}

@misc{xi2023risepotentiallargelanguage,
      title={The Rise and Potential of Large Language Model Based Agents: A Survey}, 
      author={Zhiheng Xi and Wenxiang Chen and Xin Guo and Wei He and Yiwen Ding and Boyang Hong and Ming Zhang and Junzhe Wang and Senjie Jin and Enyu Zhou and Rui Zheng and Xiaoran Fan and Xiao Wang and Limao Xiong and Yuhao Zhou and Weiran Wang and Changhao Jiang and Yicheng Zou and Xiangyang Liu and Zhangyue Yin and Shihan Dou and Rongxiang Weng and Wensen Cheng and Qi Zhang and Wenjuan Qin and Yongyan Zheng and Xipeng Qiu and Xuanjing Huang and Tao Gui},
      year={2023},
      eprint={2309.07864},
      archivePrefix={arXiv},
      primaryClass={cs.AI},
      url={https://arxiv.org/abs/2309.07864}, 
}

@Article{Torrens02012024,
  author    = {Paul M. Torrens and Ryan Kim and},
  journal   = {Annals of GIS},
  title     = {Evoking embodiment in immersive geosimulation environments},
  year      = {2024},
  number    = {1},
  pages     = {35--66},
  volume    = {30},
  doi       = {10.1080/19475683.2024.2316601},
  eprint    = {https://doi.org/10.1080/19475683.2024.2316601},
  publisher = {Taylor \& Francis},
  url       = {https://doi.org/10.1080/19475683.2024.2316601},
}

@inproceedings{10.1145/3681770.3698571,
author = {Yin, Fuzhen and Jiang, Na and Crooks, Andrew and Laurian, Lucie},
title = {Agent-based modeling of COVID-19 vaccine uptake in New York State: Information diffusion in hybrid spaces},
year = {2024},
isbn = {9798400711497},
publisher = {Association for Computing Machinery},
address = {New York, NY, USA},
url = {https://doi.org/10.1145/3681770.3698571},
doi = {10.1145/3681770.3698571},
booktitle = {Proceedings of the 7th ACM SIGSPATIAL International Workshop on GeoSpatial Simulation},
pages = {11–20},
numpages = {10},
location = {Atlanta, GA, USA},
series = {GeoSim '24}
}

@Article{Kim2020,
  author     = {Kim, Joon-Seok and Kavak, Hamdi and Rouly, Chris Ovi and Jin, Hyunjee and Crooks, Andrew and Pfoser, Dieter and Wenk, Carola and Z\"{u}fle, Andreas},
  journal    = {SIGSPATIAL Special},
  title      = {Location-Based Social Simulation for Prescriptive Analytics of Disease Spread},
  year       = {2020},
  month      = {jun},
  number     = {1},
  pages      = {53–61},
  volume     = {12},
  address    = {New York, NY, USA},
  doi        = {10.1145/3404820.3404828},
  issue_date = {March 2020},
  numpages   = {9},
  publisher  = {Association for Computing Machinery},
  url        = {https://doi.org/10.1145/3404820.3404828},
}

@InProceedings{Mekni2008,
  author    = {Mekni, Mehdi and Moulin, Bernard},
  booktitle = {Proceedings of the 5th International Conference on Soft Computing as Transdisciplinary Science and Technology},
  title     = {Using Multi-Agent Geo-Simulation Techniques for Intelligent Sensor Web Management},
  year      = {2008},
  address   = {New York, NY, USA},
  pages     = {669–674},
  publisher = {Association for Computing Machinery},
  series    = {CSTST '08},
  doi       = {10.1145/1456223.1456358},
  isbn      = {9781605580463},
  location  = {Cergy-Pontoise, France},
  numpages  = {6},
  url       = {https://doi.org/10.1145/1456223.1456358},
}

@InProceedings{Perron2008,
  author    = {Perron, Jimmy and Hogan, Jimmy and Moulin, Bernard and Berger, Jean and B\'{e}langer, Micheline},
  booktitle = {Proceedings of the 40th Conference on Winter Simulation},
  title     = {A Hybrid Approach Based on Multi-Agent Geosimulation and Reinforcement Learning to Solve a UAV Patrolling Problem},
  year      = {2008},
  address   = {Miami, Florida},
  pages     = {1259–1267},
  publisher = {Winter Simulation Conference},
  series    = {WSC '08},
  isbn      = {9781424427086},
  numpages  = {9},
}

@InProceedings{Mekni2011,
  author    = {Mekni, Mehdi and Moulin, Bernard},
  booktitle = {Proceedings of the Winter Simulation Conference},
  title     = {Informed Virtual Geographic Environments: A Geometrically Precise and Semantically Enriched Model for Multi-Agent Geo-Simulations},
  year      = {2011},
  address   = {Phoenix, Arizona},
  pages     = {381–392},
  publisher = {Winter Simulation Conference},
  series    = {WSC '11},
  numpages  = {12},
}

@Article{Wozniak2020,
  author    = {Marcin Wozniak},
  title     = {{Virtualising Space {\textendash} New Directions for Applications of Agent-Based Modelling in Spatial Economics}},
  year      = {2020},
  language  = {{EN}},
  publisher = {{Lodz University Press}},
  url       = {https://doi.org/10.18778/0208-6018.346.01},
}

@Article{Fisher2017,
  author    = {Rohan Fisher AND Jonatan Lassa},
  title     = {{Interactive, open source, travel time scenario modelling: tools to facilitate participation in health service access analysis}},
  year      = {2017},
  language  = {{EN}},
  publisher = {{BMC}},
  url       = {https://doi.org/10.1186/s12942-017-0086-8},
}

@Article{Rienow2014,
  author    = {Rienow, Andreas and Stenger, Dirk},
  journal   = {Journal of Geographical Systems},
  title     = {Geosimulation of urban growth and demographic decline in the Ruhr: a case study for 2025 using the artificial intelligence of cells and agents},
  year      = {2014},
  issn      = {1435-5930,1435-5949},
  pages     = {311--342},
  volume    = {16},
  doi       = {10.1007/s10109-014-0196-9},
  issue     = {3},
  publisher = {Springer-Verlag},
  url       = {http://doi.org/10.1007/s10109-014-0196-9},
}

@Article{Dragicevic2018,
  author  = {Dragićević, Suzana and Hatch, Kristofer},
  journal = {Habitat International},
  title   = {Urban geosimulations with the Logic Scoring of Preference method for agent-based decision-making},
  year    = {2018},
  issn    = {01973975},
  month   = feb,
  pages   = {3--17},
  volume  = {72},
  doi     = {10.1016/j.habitatint.2017.09.006},
  langid  = {english},
  url     = {https://linkinghub.elsevier.com/retrieve/pii/S0197397516307561},
  urldate = {2022-04-19},
}

@Article{Torrens2013,
  author     = {Torrens, Paul and Kevrekidis, Yannis and Ghanem, Roger and Zou, Yu},
  journal    = {Entropy},
  title      = {Simple Urban Simulation Atop Complicated Models: Multi-Scale Equation-Free Computing of Sprawl Using Geographic Automata},
  year       = {2013},
  issn       = {1099-4300},
  month      = jul,
  number     = {12},
  pages      = {2606--2634},
  volume     = {15},
  doi        = {10.3390/e15072606},
}

@InProceedings{Torrens2021,
  author     = {Torrens, Paul M. and Gu, Simin},
  booktitle  = {Proceedings of the 4th {ACM} {SIGSPATIAL} International Workshop on {GeoSpatial} Simulation},
  title      = {Real-time experiential geosimulation in virtual reality with immersion-emission},
  year       = {2021},
  address    = {Beijing China},
  month      = nov,
  pages      = {19--28},
  publisher  = {{ACM}},
  doi        = {10.1145/3486184.3491079},
}

@Article{Macatulad2014,
  author   = {Macatulad, E. G. and Blanco, A. C.},
  journal  = {Int. Arch. Photogramm. Remote Sens. Spatial Inf. Sci.},
  title    = {3DGIS-Based Multi-Agent Geosimulation and Visualization of Building Evacuation Using {GAMA} Platform},
  year     = {2014},
  issn     = {2194-9034},
  month    = nov,
  pages    = {87--91},
  volume   = {{XL}-2},
  doi      = {10.5194/isprsarchives-XL-2-87-2014},
}

@Article{Torrens2005,
  author  = {Torrens, Paul M. and Benenson, Itzhak},
  journal = {International Journal of Geographical Information Science},
  title   = {Geographic Automata Systems},
  year    = {2005},
  issn    = {1365-8816, 1362-3087},
  month   = apr,
  number  = {4},
  pages   = {385--412},
  volume  = {19},
  doi     = {10.1080/13658810512331325139},
}

@InCollection{Blecic2009,
  author    = {Blecic, Ivan and Cecchini, Arnaldo and Trunfio, Giuseppe A.},
  booktitle = {Geocomputation and Urban Planning},
  publisher = {Springer Berlin Heidelberg},
  title     = {A Multi-Agent Geosimulation Infrastructure for Planning},
  year      = {2009},
  address   = {Berlin, Heidelberg},
  editor    = {Murgante, Beniamino and Borruso, Giuseppe and Lapucci, Alessandra},
  isbn      = {978-3-540-89929-7 978-3-540-89930-3},
  note      = {{ISSN}: 1860-949X, 1860-9503 Series Title: Studies in Computational Intelligence},
  pages     = {237--253},
  volume    = {176},
  langid    = {english},
  url       = {http://link.springer.com/10.1007/978-3-540-89930-3},
  urldate   = {2022-04-19},
}

@Article{Bandyopadhyay2018,
  author  = {Bandyopadhyay, Mainak and Singh, Varun},
  journal = {Arab J Geosci},
  title   = {Agent-based geosimulation for assessment of urban emergency response plans},
  year    = {2018},
  issn    = {1866-7511, 1866-7538},
  month   = apr,
  number  = {8},
  pages   = {165},
  volume  = {11},
  doi     = {10.1007/s12517-018-3523-5},
  langid  = {english},
  url     = {http://link.springer.com/10.1007/s12517-018-3523-5},
  urldate = {2022-04-19},
}

@Article{Vahidnia2015,
  author  = {Vahidnia, Mohammad H. and Alesheikh, Ali A. and Alavipanah, Seyed Kazem},
  journal = {J Geogr Syst},
  title   = {A multi-agent architecture for geosimulation of moving agents},
  year    = {2015},
  issn    = {1435-5930, 1435-5949},
  month   = oct,
  number  = {4},
  pages   = {353--390},
  volume  = {17},
  doi     = {10.1007/s10109-015-0218-2},
  langid  = {english},
  url     = {http://link.springer.com/10.1007/s10109-015-0218-2},
  urldate = {2022-04-19},
}

@InProceedings{VanegasHernandez2017,
  author     = {Vanegas-Hernandez, Meili and da Costa Pereira, Célia and Moreno, Diego and Fusco, Giovanni and Tettamanzi, Andrea G. B. and Riveill, Michel and Hernández, José Tiberio},
  booktitle  = {Proceedings of the International Conference on Web Intelligence},
  title      = {A new urban segregation-growth coupled model using a belief-desire-intention possibilistic framework},
  year       = {2017},
  address    = {Leipzig Germany},
  month      = aug,
  pages      = {340--347},
  publisher  = {{ACM}},
  doi        = {10.1145/3106426.3106486},
  eventtitle = {{WI} '17: International Conference on Web Intelligence 2017},
  isbn       = {978-1-4503-4951-2},
  langid     = {english},
  url        = {https://dl.acm.org/doi/10.1145/3106426.3106486},
  urldate    = {2022-04-19},
}

@Article{Pires2017,
  author     = {Pires, Bianica and Crooks, Andrew T.},
  journal    = {Computers, Environment and Urban Systems},
  title      = {Modeling the emergence of riots: A geosimulation approach},
  year       = {2017},
  issn       = {01989715},
  month      = jan,
  pages      = {66--80},
  volume     = {61},
  doi        = {10.1016/j.compenvurbsys.2016.09.003},
  langid     = {english},
  shorttitle = {Modeling the emergence of riots},
  url        = {https://linkinghub.elsevier.com/retrieve/pii/S0198971516302459},
  urldate    = {2022-04-19},
}

@Article{Raimbault2020,
  author  = {Raimbault, Juste and Broere, Joris and Somveille, Marius and Serna, Jesus Mario and Strombom, Evelyn and Moore, Christine and Zhu, Ben and Sugar, Lorraine},
  journal = {Resources, Conservation and Recycling},
  title   = {A spatial agent based model for simulating and optimizing networked eco-industrial systems},
  year    = {2020},
  issn    = {09213449},
  month   = apr,
  pages   = {104538},
  volume  = {155},
  doi     = {10.1016/j.resconrec.2019.104538},
  langid  = {english},
  url     = {https://linkinghub.elsevier.com/retrieve/pii/S0921344919304446},
  urldate = {2022-04-19},
}

@Article{Macatulad2019,
  author  = {Macatulad, E. G. and Blanco, A. C.},
  journal = {International Journal of Urban Sciences},
  title   = {A 3DGIS multi-agent geo-simulation model for assessment of building evacuation scenarios considering urgency and knowledge of exits},
  year    = {2019},
  issn    = {1226-5934, 2161-6779},
  month   = jul,
  number  = {3},
  pages   = {318--334},
  volume  = {23},
  doi     = {10.1080/12265934.2018.1549505},
  langid  = {english},
  url     = {https://www.tandfonline.com/doi/full/10.1080/12265934.2018.1549505},
  urldate = {2022-04-19},
}

@Article{Haddad2019,
  author  = {Haddad, Hedi and Bouyahia, Zied and Chaudhry, Shafique A.},
  journal = {Procedia Computer Science},
  title   = {A Multiagent Geosimulation and {IoT}-based Framework for Safety Monitoring in Complex Dynamic Spatial Environments},
  year    = {2019},
  issn    = {18770509},
  pages   = {527--534},
  volume  = {151},
  doi     = {10.1016/j.procs.2019.04.071},
  langid  = {english},
  url     = {https://linkinghub.elsevier.com/retrieve/pii/S1877050919305332},
  urldate = {2022-04-19},
}

@InProceedings{Cabs2013,
  author    = {Cabs, Kaspars and Lektauers, Arnis and Merkuryev, Yuri},
  booktitle = {2013 8th EUROSIM Congress on Modelling and Simulation},
  title     = {Automated Geosimulation Approach to Urban Territory Development Planning},
  year      = {2013},
  month     = {Sep.},
  pages     = {318-323},
  doi       = {10.1109/EUROSIM.2013.64},
}

@InProceedings{Zuefle2020,
  author    = {Züfle, Andreas and Trajcevski, Goce and Pfoser, Dieter and Kim, Joon-Seok},
  booktitle = {2020 21st IEEE International Conference on Mobile Data Management (MDM)},
  title     = {Managing Uncertainty in Evolving Geo-Spatial Data},
  year      = {2020},
  month     = {June},
  pages     = {5-8},
  doi       = {10.1109/MDM48529.2020.00021},
  issn      = {2375-0324},
}

@Article{Benenson2004,
  author    = {Itzhak Benenson and Paul M.Torrens},
  journal   = {Computers, Environment and Urban Systems},
  title     = {Geosimulation: object-based modeling of urban phenomena},
  year      = {2004},
  pages     = {1-8},
  volume    = {28},
  url       = {www.elsevier.com/locate/compenvurbsys},
}

@InCollection{Bradshaw1997,
  author    = {Jeffrey M. Bradshaw},
  booktitle = {Software Agents},
  publisher = {AAAI Press / The MIT Press},
  title     = {An Introduction to Software Agents},
  year      = {1997},
  editor    = {Jeffrey M. Bradshaw},
  pages     = {3--46},
  url       = {citeseer.ist.psu.edu/bradshaw97introduction.html},
}

@Book{Bratman1987,
  author    = {M Bratman},
  editor    = {Harvard University Press},
  publisher = {Harvard University Press},
  title     = {Intention, Plans and Practical Reasoning},
  year      = {1987},
  address   = {Cambridge.},
}

@Article{Bresciani2004,
  author    = {P Bresciani and P Giorgini and F Giunchiglia and J Mylopoulos and A Perini.},
  journal   = {Autonomous Agents and Multi-Agent Systems,},
  title     = {Tropos: An Agent-Oriented Software Development Methodology.},
  year      = {2004},
  number    = {No. 3},
  pages     = {pp. 203--236},
  volume    = {Vol. 8, no.3},
  booktitle = {Journal of Autonomous Agents and Multi-Agent Systems},
  url       = {citeseer.ist.psu.edu/article/bresciani04tropos.html},
}

@Article{Caire2002,
  author    = {G Caire and W Coulier and F Garijo and J Gomez and J Pavon and F Leal and P Chainho and P E Kearney and J Stark and R Evans and P Massonet},
  journal   = {LNCS},
  title     = {Agent Oriented Analysis Using Message/{UML}.},
  year      = {2002},
  pages     = {pp. 119-135},
  volume    = {2222},
  booktitle = {AOSE},
  url       = {citeseer.ist.psu.edu/caire01agent.html},
}

@PhdThesis{Davila1997,
  author    = {Jacinto D\'avila},
  school    = {Imperial College of Science, Technology and Medicine},
  title     = {Agents in Logic Programming},
  year      = {1997},
  address   = {London, UK},
  month     = {June},
}

@Article{Davila2005,
  author    = {Jacinto D\'avila and E G\'{o}mez and K Laffaille and K Tucci and M Uzc\'{a}tegui.},
  journal   = {The 9-th IEEE International Symposium on Distributed Simulation and Real Time Applications},
  title     = {Multi-Agent Distributed Simulations with GALATEA.},
  year      = {2005},
  note      = {ISBN 0-7695-2462-1},
  pages     = {pp. 165-170},
  address   = {Montreal, Canada},
  booktitle = {DS-RT'2005. The 9-th IEEE International Symposium on Distributed Simulation and Real Time Applications},
  editor    = {In A. Boukerche, S. Turner, D. Roberts and G. Theodoropoulos (eds).},
  series    = {IEEE Computer Society},
}

@Article{Davila2002,
  author    = {Jacinto D\'avila and Kay Tucci},
  journal   = {AMSE Special Issue 2000. Association for the advancement of Modelling \& Simulation techniques in Enterprises},
  title     = {Towards a logic-based, multi-agent simulation theory},
  year      = {2002},
  note      = {Lion, France},
  pages     = {37-51},
}

@Article{Davila2005b,
  author    = {Jacinto D\'avila and Mayerl\'in Uzc\'ategui},
  journal   = {The Fifth IASTED Internacional Conference on Modelling, Simulation and Optimization, Oranjestad, Aruba},
  title     = {Agents that learn to behave in Multi-Agent Simulations},
  year      = {2005},
  month     = {August},
  pages     = {pp. 51-55},
  address   = {Oranjestad, Aruba},
  booktitle = {MSO'2005. Fifth IASTED International Conference on Modelling, Simulation and Optimization},
}

@Article{Davila2002a,
  author    = {Jacinto D\'avila and Mayerl\'in Uzc\'ategui},
  journal   = {AMSE Special Issue 2000. Association for the advancement of Modelling \& Simulation techniques in Enterprises},
  title     = {GALATEA: A multi-agent simulation platform},
  year      = {2002},
  note      = {Lion, France},
  pages     = {52-67},
}

@InProceedings{Davila2000,
  author    = {Jacinto D\'avila and Mayerlin Uzc\'ategui},
  booktitle = {Proceedings of MSNN2000: International Conference on Modeling, Simulation and Neural Networks. M\'erida. Venezuela.},
  title     = {{GALATEA}: A Multi-agent Simulation Platform.},
  year      = {2000},
}

@InProceedings{Davila2007,
  author    = {Jacinto D\'avila and Mayerlin Uzc\'ategui and Kay Tucci},
  booktitle = {Proceedings of 2007 Summer Computer Simulation Conference (SCSC0'7). The Society For Modelling and Simulation International.},
  title     = {From a Multi-Agent Simulation Theory to Galatea.},
  year      = {2007},
}

@InProceedings{Davila2005a,
  author    = {Jacinto D\'avila and Mayerlin Uzc\'ategui and Kay Tucci},
  booktitle = {MSO'2005. Fifth IASTED International Conference on Modelling, Simulation and Optimization},
  title     = {A Multi-Agent Theory for Simulation.},
  year      = {2005},
  address   = {Oranjestad, Aruba},
  month     = {August},
  pages     = {285-290},
  journal   = {The Fifth IASTED Internacional Conference on Modelling, Simulation and Optimization (MSO 2005), Oranjestad, Aruba.},
}

@Article{Ferber1996,
  author    = {J Ferber and J-P Muller},
  journal   = {Second International Conference on MultiAgent Systems},
  title     = {Influences and Reaction: a Model of Situated Multiagent Systems.},
  year      = {1996},
  pages     = {pp. 72-79},
  booktitle = {ICMAS-96. Second International Conference on Multiagent Systems},
  url       = {http://www.aaai.org/Papers/ICMAS/1996/ICMAS96-009.pdf},
}

@Article{GomezSanz2003,
  author    = {JJ G\'{o}mez-Sanz and J Pav\'{o}n},
  title     = {Agent Oriented Software Engineering with INGENIAS.},
  year      = {2003},
  pages     = {pp. 394-403},
  volume    = {Vol. 2691 of LNCS},
  address   = {Prague, Czech Republic},
  booktitle = {3er International Central and Eastern European Conference on MultiAgent System. CEEMAS},
  editor    = {Y. Marik and J. M\"{u}ller and M. Pechoucek},
  series    = {Multi Agent Systems and Aplications III},
  url       = {http://citeseerx.ist.psu.edu/viewdoc/summary?doi=10.1.1.111.4611},
}

@InProceedings{GarciaOjeda2007,
  author    = {Juan C. Garcia-Ojeda and Scott A. DeLoach and Robby and Walamitien H. Oyenan and Jorge Valenzuela},
  booktitle = {Proceddings of the 8th International Workshop on Agent Oriented Software Enginnering},
  title     = {O-MaSE: A Customizable Approach to Developing Multiagent Development Processes},
  year      = {2007},
  address   = {Honolulu HI},
  month     = {May},
}

@InProceedings{Iglesias1996,
  author    = {Iglesias, C and Garijo, M and Gonzalez, J and Velasco, J.R.},
  title     = {A Methodological Proposal for Multiagent Systems Development Extending CommonKADS},
  year      = {1996},
  url       = {http://citeseerx.ist.psu.edu/viewdoc/summary?doi=10.1.1.54.4868},
}

@Book{Kowalski2011,
  author    = {Robert Kowalski},
  publisher = {Cambridge University Press},
  title     = {Computational Logic and Human Thinking: How to be Artificially Intelligent},
  year      = {2011},
}

@Article{Kowalski2006,
  author    = {Robert Kowalski},
  journal   = {In: Reasoning, Action and Interaction in AI Theories and Systems - Festschrift in Honor of Luigia Carlucci Aiello. (eds. O. Stock, M. Schaerf) Springer Verlag, LNAI},
  title     = {Computational Logic in an Object-Oriented World},
  year      = {2006},
}

@Article{Kowalski2009,
  author    = {Robert Kowalski and Fariba Sadri},
  journal   = {In Web Reasoning and Rule Systems (eds. A. Polleres and T. Swift) Springer, LNCS 5837},
  title     = {Integrating Logic Programming and Production Systems in Abductive Logic Programming Agents},
  year      = {2009},
}

@Article{Kowalski1999,
  author    = {Robert Kowalski and Fariba Sadri},
  journal   = {In: Annals of Mathematics and Artificial Intelligence},
  title     = {From Logic Programming towards Multi-agent Systems ,},
  year      = {1999},
  pages     = {391-419},
  volume    = {25},
}

@Article{Kowalski1997,
  author    = {Robert Kowalski and Fariba Sadri},
  journal   = {Department of Computing, Imperial College},
  title     = {An Agent Architecture that Unifies Rationality with Reactivity},
  year      = {1997},
  url       = {http://www.doc.ic.ac.uk/~rak/},
}

@Article{Kowalski1987,
  author    = {Robert Kowalski and Fariba Sadri and Soper, P},
  journal   = {In Proceedings of VLDB, Morgan Kaufmann, Los Altos, Ca.},
  title     = {Integrity Checking in Deductive Databases},
  year      = {1987},
  pages     = {61-69},
}

@TechReport{Mark2000,
  author    = {W. Mark},
  title     = {Multiagent Systems Engineering: A Methodology for Analysis and Design of Multiagent Systems},
  year      = {2000},
  note      = {http://macr.cis.ksu.edu/projects/mase.htm},
  school    = {MS thesis, AFIT/GCS/ENG/00M-26. School of Engineering, Air Force Institute of Technology (AU), Wright-Patterson AFB, OH.},
  url       = {citeseer.ist.psu.edu/mark00multiagent.html},
}

@InProceedings{Padgham2002,
  author    = {L Padgham and M Winikoff},
  booktitle = {In Proceedings of the OOPSLA 2002 Workshop on Agent-Oriented Methodologies},
  title     = {Prometheus: A Pragmatic Methodology for Engineering Intelligent Agents.},
  year      = {2002},
  pages     = {pp. 97--108},
  url       = {http://citeseer.ist.psu.edu/article/padgham02prometheus.html},
}

@Article{Padilla2012,
  author    = {Virginia Padilla and Jacinto D\'avila},
  journal   = {LPAR-18},
  title     = {A reference model and a theory for multiagent, information systems},
  year      = {2012},
}

@PhdThesis{padilla2013a,
  Title                    = {Modelo de Referencia para Geosimulaci{\'o}n con Agentes y Bases de Datos},
  Author                   = {Padilla Sifontes, Virginia},
  School                   = {Universidad de Los Andes},
  Year                     = {2013},
  Note                     = {Doctorado en Ciencias Aplicadas a la Ingenier{\'\i}a},
  Url			   = {http://webdelprofesor.ula.ve/ingenieria/jacinto/tesis/phd-virginia-padilla.pdf} 
}

@TechReport{Rao1995,
  author    = {Anand S. Rao and Michael P. Georgeff},
  title     = {Formal Models and Decision Procedures for Multi-Agent Systems},
  year      = {1995},
  url       = {http://citeseerx.ist.psu.edu/viewdoc/summary?doi=10.1.1.52.7924},
}

@InProceedings{Rao1995a,
  author    = {Anand S. Rao and Michael P. Georgeff},
  booktitle = {Proceedings of the First Intl. Conference on Multiagent Systems},
  title     = {BDI-agents: from theory to practice},
  year      = {1995},
  address   = {San Francisco},
  url       = {citeseer.ist.psu.edu/rao95bdi.html},
}

@Article{Rao1991,
  author    = {Anand S. Rao and Michael P. Georgeff},
  journal   = {Second International Conference on Principles of Knowledge Representation and Reasoning.},
  title     = {Modeling rational agents within a BDI-architecture},
  year      = {1991},
  url       = {http://citeseerx.ist.psu.edu/viewdoc/summary?doi=10.1.1.41.3036},
}

@Book{Rumbaugh1991,
  author    = {James Rumbaugh and Michael Blaha and William Premerlani and Frederick Eddy and William Lorensen},
  publisher = {Prentice Hall},
  title     = {Object-Oriented Modeling and Design},
  year      = {1991},
  isbn      = {ISBN 0-13-629841-9},
}

@Book{Russell2004,
  author    = {S Russell and P Norvig},
  editor    = {Pearson},
  publisher = {Pearson},
  title     = {Inteligencia Artificial: Un enfoque moderno. 7ma edici\'{o}n.},
  year      = {2004},
  address   = {Espa\~na},
}

@Article{Shoham1991,
  author    = {Y Shoham},
  journal   = {AAAI-91 Proceedings},
  title     = {{AGENT}{0}: A Simple Agent Language and Its Interpreter.},
  year      = {1991},
  number    = {No. 91},
  pages     = {pp. 704-709},
  bibsource = {DBLP, http://dblp.uni-trier.de},
  booktitle = {AAAI},
}

@Book{Wainer2009,
  author    = {Gabriel Wainer},
  editor    = {CRC Press},
  title     = {Discrete-Event Modeling and Simulation},
  year      = {2009},
}

@InProceedings{Wooldridge1998,
  author    = {Michael Wooldridge and Nicholas R. Jennings},
  booktitle = {Proceedings of the 2nd International Conference on Autonomous Agents (Agents'98)},
  title     = {Pitfalls of Agent-Oriented Development},
  year      = {1998},
  address   = {New York},
  editor    = {Katia P. Sycara and Michael Wooldridge},
  pages     = {385--391},
  publisher = {ACM Press},
  isbn      = {0-89791-983-1},
  url       = {citeseer.ist.psu.edu/wooldridge98pitfalls.html},
}

@InProceedings{Zambonelli2003,
  author    = {F. Zambonelli and N. Jennings and M. Wooldridge},
  booktitle = {ACM Transactions on Software Engineering and Methodology, 12(3).},
  title     = {Developing Multiagent Systems: the Gaia Methodology},
  year      = {2003},
  url       = {citeseer.ist.psu.edu/zambonelli03developing.html},
}

@Book{Zeigler2000,
  author    = {Bernard P. Zeigler and Herbert Praehofer and Tag Gon Kim},
  publisher = {Academic Press},
  title     = {Theory of Modelling and Simulation},
  year      = {2000},
  edition   = {Second},
}

@Misc{Sahli2008,
  author      = {Nabil Sahli AND Mehdi Mekni AND Bernard Moulin},
  title       = {{A Multi-Agent Geo-Simulation Approach for the Identification of Risky Areas for Trains}},
  year        = {2008},
  contributor = {{The Pennsylvania State University CiteSeerX Archives}},
  language    = {{en}},
  url         = {http://citeseerx.ist.psu.edu/viewdoc/summary?doi=10.1.1.208.9783},
}

@InProceedings{BlumenfeldLieberthal2012,
  author    = {Blumenfeld-Lieberthal, Efrat and Portugali, Juval},
  booktitle = {Self-Organizing Systems},
  title     = {MetroNet: A Metropolitan Simulation Model Based on Commuting Processes},
  year      = {2012},
  address   = {Berlin, Heidelberg},
  editor    = {Kuipers, Fernando A. and Heegaard, Poul E.},
  pages     = {96--103},
  publisher = {Springer Berlin Heidelberg},
  isbn      = {978-3-642-28583-7},
}

@Article{Fecht2014,
  author    = {Fecht, Daniela and Beale, Linda and Briggs, David},
  journal   = {Environmental Modelling \& Software},
  title     = {A GIS-based urban simulation model for environmental health analysis},
  year      = {2014},
  issn      = {1364-8152},
  pages     = {1--11},
  volume    = {58},
  doi       = {10.1016/j.envsoft.2014.03.013},
  publisher = {Elsevier Science},
  url       = {http://doi.org/10.1016/j.envsoft.2014.03.013},
}

@InProceedings{Fu2020,
  author    = {Fu, Yanyun and Zeng, Yiping and Wang, Deyong and Zhang, Hui and Gao, Yang and Liu, Yi},
  booktitle = {2020 {International} {Conference} on {Urban} {Engineering} and {Management} {Science} ({ICUEMS})},
  title     = {Research on {Route} {Optimization} {Based} on {Multiagent} and {Genetic} {Algorithm} for {Community} {Patrol}},
  year      = {2020},
  month     = apr,
  pages     = {112--116},
  doi       = {10.1109/ICUEMS50872.2020.00034},
}

@InProceedings{Kim2019a,
  author    = {Kim, Joon-Seok and Kavak, Hamdi and Manzoor, Umar and Crooks, Andrew and Pfoser, Dieter and Wenk, Carola and Z\"{u}fle, Andreas},
  booktitle = {Proceedings of the 27th ACM SIGSPATIAL International Conference on Advances in Geographic Information Systems},
  title     = {Simulating Urban Patterns of Life: A Geo-Social Data Generation Framework},
  year      = {2019},
  address   = {New York, NY, USA},
  pages     = {576–579},
  publisher = {Association for Computing Machinery},
  series    = {SIGSPATIAL '19},
  doi       = {10.1145/3347146.3359106},
  isbn      = {9781450369091},
  location  = {Chicago, IL, USA},
  numpages  = {4},
  url       = {https://doi.org/10.1145/3347146.3359106},
}

@Article{Sahli2009,
  author    = {Nabil Sahli and Bernard Moulin},
  journal   = {Applied Intelligence},
  title     = {EKEMAS, an agent-based geo-simulation framework to support continual planning in the real-word},
  year      = {2009},
  issn      = {0924-669X,1573-7497},
  pages     = {188--209},
  volume    = {31},
  doi       = {10.1007/s10489-008-0122-2},
  issue     = {2},
  publisher = {Springer US},
  url       = {http://doi.org/10.1007/s10489-008-0122-2},
}

@Article{Schelling1971,
  author    = {Thomas C. Schelling},
  journal   = {The Journal of Mathematical Sociology},
  title     = {Dynamic models of segregation},
  year      = {1971},
  number    = {2},
  pages     = {143-186},
  volume    = {1},
  doi       = {10.1080/0022250X.1971.9989794},
  eprint    = {https://doi.org/10.1080/0022250X.1971.9989794},
  publisher = {Routledge},
  url       = {https://doi.org/10.1080/0022250X.1971.9989794},
}

@InProceedings{Moulin2003,
  author    = {Moulin, Bernard and Chaker, Walid and Perron, Jimmy and Pelletier, Patrick and Hogan, Jimmy and Gbei, Edouard},
  booktitle = {Spatial Information Theory. Foundations of Geographic Information Science},
  title     = {MAGS Project: Multi-agent GeoSimulation and Crowd Simulation},
  year      = {2003},
  address   = {Berlin, Heidelberg},
  editor    = {Kuhn, Walter and Worboys, Michael F. and Timpf, Sabine},
  pages     = {151--168},
  publisher = {Springer Berlin Heidelberg},
  isbn      = {978-3-540-39923-0},
}

@Article{Maslow1943,
  author  = {Maslow, A. H.},
  journal = {Psychological Review,},
  title   = {A theory of human motivation.},
  year    = {1943},
  number  = {4},
  pages   = {370–396},
  volume  = {50},
  doi     = {https://doi.org/10.1037/h0054346},
}

@Article{Sullivan2010,
  author = {Sullivan, Keith and Coletti, Mark and Luke, Sean},
  title  = {GeoMason: Geospatial Support for MASON},
  year   = {2010},
  month  = {06},
}

@Article{Bergenti2014,
  author  = {Bergenti, Federico and Caire, Giovanni and Gotta, Danilo},
  journal = {CEUR Workshop Proceedings},
  title   = {Agents on the move: JADE for android devices},
  year    = {2014},
  month   = {01},
  volume  = {1260},
}

@InProceedings{CostaPereira2010,
  author  = {da Costa Pereira, Celia and Tettamanzi, Andrea},
  title   = {An integrated possibilistic framework for goal generation in cognitive agents},
  year    = {2010},
  month   = {01},
  pages   = {1239-1246},
  volume  = {2},
  doi     = {10.1145/1838206.1838367},
  journal = {Proceedings of the International Joint Conference on Autonomous Agents and Multiagent Systems, AAMAS},
}

@Article{Haddad2010,
  author    = {Haddad, Hedi; Moulin, Bernard},
  journal   = {Journal of Experimental \& Theoretical Artificial Intelligence},
  title     = {A framework to support qualitative reasoning about COAs in a dynamic spatial environment},
  year      = {2010},
  issn      = {0952-813X,1362-3079},
  pages     = {341--380},
  volume    = {22},
  doi       = {10.1080/09528131003713002},
  issue     = {4},
  publisher = {Taylor and Francis Group},
  url       = {http://doi.org/10.1080/09528131003713002},
}

@Article{Benenson2008,
  author    = {Itzhak Benenson and Karel Martens and Slava Birfir},
  journal   = {Computers, Environment and Urban Systems},
  title     = {PARKAGENT: An agent-based model of parking in the city},
  year      = {2008},
  issn      = {0198-9715},
  pages     = {431--439},
  volume    = {32},
  doi       = {10.1016/j.compenvurbsys.2008.09.011},
  issue     = {6},
  publisher = {Elsevier Science},
  url       = {http://doi.org/10.1016/j.compenvurbsys.2008.09.011},
}

@InProceedings{Taillandier2012,
  author      = {Taillandier, Patrick and Vo, Duc-An and Amouroux, Edouard and Drogoul, Alexis},
  booktitle   = {{PRINCIPLES AND PRACTICE OF MULTI-AGENT SYSTEMS}},
  title       = {{GAMA : a simulation platform that integrates geographical information data, agent-based modeling and multi-scale control}},
  year        = {2012},
  address     = {Kolkata, India},
  pages       = {242-258},
  publisher   = {{Springer Berlin / Heidelberg}},
  series      = {Lecture Notes in Computer Science},
  volume      = {7057/2012},
  doi         = {10.1007/978-3-642-25920-3\_17},
  hal_id      = {hal-00688318},
  hal_version = {v2},
  url         = {https://hal.science/hal-00688318},
}

@InBook{Kennedy2012,
  author    = {Kennedy, William G.},
  editor    = {Heppenstall, Alison J. and Crooks, Andrew T. and See, Linda M. and Batty, Michael},
  pages     = {167--179},
  publisher = {Springer Netherlands},
  title     = {Modelling Human Behaviour in Agent-Based Models},
  year      = {2012},
  address   = {Dordrecht},
  isbn      = {978-90-481-8927-4},
  booktitle = {Agent-Based Models of Geographical Systems},
  doi       = {10.1007/978-90-481-8927-4},
  url       = {https://doi.org/10.1007/978-90-481-8927-4_9},
}

@Article{Williams2011,
  author   = {Williams, Richard J and Purves, Drew W},
  journal  = {Ecology},
  title    = {The probabilistic niche model reveals substantial variation in the niche structure of empirical food webs},
  year     = {2011},
  number   = {9},
  pages    = {1849-1857},
  volume   = {92},
  doi      = {https://doi.org/10.1890/11-0200.1},
  eprint   = {https://esajournals.onlinelibrary.wiley.com/doi/pdf/10.1890/11-0200.1},
  url      = {https://esajournals.onlinelibrary.wiley.com/doi/abs/10.1890/11-0200.1},
}

@Article{Sabri2012,
  author   = {Sabri, Soheil and M. Ludin, Ahmad Nazri Muhammad and Ho, Chin Siong},
  journal  = {Applied Spatial Analysis and Policy},
  title    = {Conceptual {Design} for an {Integrated} {Geosimulation} and {Analytic} {Network} {Process} ({ANP}) in {Gentrification} {Appraisal}},
  year     = {2012},
  issn     = {1874-4621},
  month    = sep,
  number   = {3},
  pages    = {253--271},
  volume   = {5},
  doi      = {10.1007/s12061-011-9069-5},
  url      = {https://doi.org/10.1007/s12061-011-9069-5},
}

@Article{Epstein2002,
  author    = {Epstein, J. M.},
  journal   = {Proceedings of the National Academy of Sciences},
  title     = {Modeling civil violence: An agent-based computational approach},
  year      = {2002},
  issn      = {0027-8424,1091-6490},
  number    = {Supplement 3},
  pages     = {7243--7250},
  volume    = {99},
  publisher = {National Academy of Sciences},
  url       = {http://doi.org/10.1073/pnas.092080199},
}

@Article{Torrens2015a,
  author  = {Torrens, Paul},
  journal = {Annals of GIS},
  title   = {Slipstreaming human geosimulation in virtual geographic environments},
  year    = {2015},
  month   = {02},
  pages   = {1-20},
  volume  = {21},
  doi     = {10.1080/19475683.2015.1009489},
}

@Article{MahdizadehGharakhanlou2022,
  author         = {Mahdizadeh Gharakhanlou, Navid and Perez, Liliana},
  journal        = {ISPRS International Journal of Geo-Information},
  title          = {Geocomputational Approach to Simulate and Understand the Spatial Dynamics of COVID-19 Spread in the City of Montreal, QC, Canada},
  year           = {2022},
  issn           = {2220-9964},
  number         = {12},
  volume         = {11},
  article-number = {596},
  doi            = {10.3390/ijgi11120596},
  url            = {https://www.mdpi.com/2220-9964/11/12/596},
}

@Article{Gibson1979,
  author  = {Gibson, J. J.},
  journal = {Houghton, Mifflin and Company.},
  title   = {The ecological approach to visual perception},
  year    = {1979},
}

@InProceedings{Mekni2010b,
  author    = {Mekni, Mehdi and Moulin, Bernard},
  booktitle = {2010 Second International Conference on Advanced Geographic Information Systems, Applications, and Services},
  title     = {Motion Planning of Autonomous Agents Situated in Informed Virtual Geographic Environments},
  year      = {2010},
  pages     = {1-8},
  url       = {10.1109/GEOProcessing.2010.8},
}

@InBook{Bouden2010,
  author = {Bouden, Mondher},
  title  = {Multi-Level Geosimulation of Zoonosis Propagation: a Multi-Agent and Climate Sensitive Tool for Risk Management in Public Health},
  year   = {2010},
  isbn   = {978-953-307-138-1},
  month  = {08},
  doi    = {10.5772/9906},
}

@Article{BenDor2020,
  author  = {Ben-Dor, Golan and Ben-Elia, Eran and Benenson, Itzhak},
  journal = {Procedia Computer Science},
  title   = {Spatiotemporal Implications of Population Downscaling: A MATSim Study of Sioux Falls Morning Peak Traffic},
  year    = {2020},
  month   = {01},
  pages   = {720-725},
  volume  = {170},
  doi     = {10.1016/j.procs.2020.03.165},
}

@InProceedings{masad2015,
  author       = {Masad, David and Kazil, Jacqueline and others},
  booktitle    = {14th PYTHON in Science Conference},
  title        = {MESA: an agent-based modeling framework},
  year         = {2015},
  organization = {Citeseer},
  pages        = {53--60},
  volume       = {2015},
}

@Article{Torrens2013a,
  author    = {Torrens, Paul M. and McDaniel, Aaron W.},
  journal   = {Annals of the Association of American Geographers},
  title     = {Modeling Geographic Behavior in Riotous Crowds},
  year      = {2013},
  issn      = {0004-5608,1467-8306},
  pages     = {20--46},
  volume    = {103},
  doi       = {10.1080/00045608.2012.685047},
  issue     = {1},
  publisher = {Taylor and Francis Group},
  url       = {http://doi.org/10.1080/00045608.2012.685047},
}

@Article{Luke2005,
  author  = {S. Luke and C. Cioffi-Revilla and L. Panait and K. Sullivan and G. Balan.},
  journal = {Simulation},
  title   = {A multiagent simulation environment.},
  year    = {2005},
  number  = {7},
  pages   = {517–527},
  volume  = {8},
}

@InProceedings{Padilla2013,
  author  = {Padilla, Virginia and Dávila Quintero, Jacinto},
  title   = {Multi-agent geosimulation for a water distribution System},
  year    = {2013},
  month   = {10},
  pages   = {1-12},
  doi     = {10.1109/CLEI.2013.6670651},
  isbn    = {978-1-4799-2957-3},
  journal = {Proceedings of the 2013 39th Latin American Computing Conference, CLEI 2013},
}
 
\end{document}